\documentclass[twocolumn,trackchanges]{aastex62} 

\usepackage{color}
\graphicspath{{./}{figures/}}

\received{???}
\revised{???}
\accepted{???}
\shorttitle{Sagittarius core properties}
\shortauthors{Wang et al.}

\begin{document}
\title{The Dark Matter Tidal Stripping History of the Sagittarius Core with N-body simulations}
\author[0000-0001-8459-1036]{Hai-Feng Wang}
\affil{GEPI, Observatoire de Paris, Universit\'e PSL, CNRS, Place Jules Janssen 92195, Meudon, France}
\author{Francois Hammer}
\affil{GEPI, Observatoire de Paris, Universit\'e PSL, CNRS, Place Jules Janssen 92195, Meudon, France}
\author{Yan-Bin Yang}
\affil{GEPI, Observatoire de Paris, Universit\'e PSL, CNRS, Place Jules Janssen 92195, Meudon, France}
\author{Jian-Ling Wang}
\affil{CAS Key Laboratory of Optical Astronomy, National Astronomical Observatories, Beijing 100101, China}

\correspondingauthor{HFW; FH; YBY}
\email{haifeng.wang@obspm.fr};
\email{francois.hammer@obspm.fr};
\email{yanbin.yang@obspm.fr};

\begin{abstract}

The infall of the Sagittarius (Sgr) Dwarf Spheroidal Galaxy in the Milky Way halo is an unique opportunity to understand how the different components of a dwarf galaxy could be tidally removed. In this work, we reconstruct the Sgr core morphology and kinematics on the basis of a model that has already successfully reproduced the Sgr stream. Here we use a very high resolution model  that almost resolves individual stars in the Sgr core. It reproduces most of the observed morphology and kinematic properties, without specific fine-tuning. We also show that the dark matter may have been almost entirely stripped by Milky Way tides after two passages at pericenter. Finally the model predicts that the Sgr core will be fully disrupted within the next 2 Gyr.

\end{abstract}


\section{Introduction}

Our understanding of the formation and evolution of massive galaxies has been limited for a while by the question of disk formation \citep{White1978,Navarro1995}. This has substantially progressed in the last decade by the discovery that many massive galactic disks are rejuvenated by major mergers \citep{Hammer2005,Hammer2009,Robertson2008,Stewart2009}. However, the formation and evolution of dwarf galaxies is still a matter of debate. For example, the proximity of MW dwarfs to their pericenters \citep{Simon2019,Hammer2020} may question their dark matter content. Dwarf galaxies may reveal important properties of their matter content when they are observed in a merging process such as that experienced by the Sgr dwarf galaxy.  Since this ongoing minor merger event was discovered by \citet{Ibata1994}, Sagittarius core or remnant has been investigated by many studies \citep{Ibata1995,Lokas2010,Penarrubia2010,Vasiliev2020,Vasiliev2021,delPino2021}. This dwarf galaxy may help to probe the Milky Way potential and may have affected the Milky Way disc \citep{Antoja2018,wang2018a, wang2018b, wang2019, wang2020a, wang2020b, wang2020c, Lopez2020, Yu2021,Joss2021,Yang2022, 2018MNRAS.481..286L, 2019MNRAS.485.3134L, 2020MNRAS.492L..61L}. There are also globular clusters associated to the Sgr system  such as M 54, Ter 8, Ter 7, Arp 2 inside or near the main body of the Sagittarius dwarf spheroidal galaxy, and Pal 12, Whiting 1 that belong to the trailing arm \citep{Bellazzini2020}.

Using N-body simulations under the framework of tidal stripping scenario, \citet{Lokas2010} presented the first model for the evolution of the Sgr dwarf considering the observed elliptical shape, and suggested that the total mass nowadays within 5 kpc is $5.2 \times 10^{8}M_{\odot}$ with small intrinsic rotation left. \citet{delPino2021} showed that the Sgr core has a S-shape morphology and shows a mild rotation. This has been done on the basis of a machine learning method for Gaia DR2 RR Lyrae stars and of a N-body model whose progenitor is a flattened rotating disk. However, the model is based on a Sgr-like dwarf and a relatively massive Milky Way, which could be quite different from the real Sgr orbiting around the real Milky Way. 

Many properties of the Sgr remnant or core are still under debate, e.g., the Sgr core mass and its dynamical state, which is likely affected by MW tides. \citet{Vasiliev2020} showed that a tidally disrupted Sgr making 2.5 orbits around the Milky Way could reproduce well the core observational properties, for which they found a present-day core mass of $4.0 \times 10^{8}M_{\odot}$ within a 5 kpc radius. Recently, \citet{wang2022} considered a Milky Way mass of $5.2 \times 10^{11}M_{\odot}$  and an initial Sgr mass of $9.3 \times 10^{8}M_{\odot}$ with a similar initial orbit than that used by \citet{Vasiliev2021} and for which they assumed a total MW mass of $9 \times 10^{11}M_{\odot}$.  To retrieve the Sgr stream, \citet{wang2022} have scaled down many mass properties of \citet{Vasiliev2020,Vasiliev2021}, and adopt a rotating disk for the Sgr progenitor, together with a dark matter scale length of 1.6 kpc. 

They have also been able to reproduce as well 3D spatial features of Sgr stream, including its leading and trailing arms. Moreover and conversely to the \citet{Vasiliev2020} model, \citet{wang2022} model is able to reproduce the observed  north and south bifurcations, which were firstly discovered in \citet{Belokurov2006} and \citet{2012ApJ...750...80K}, respectively (see also \citealt{2022ApJ...932L..14O}). 
 
Here we follow the \citet{wang2022} study, using a much higher resolution version of their model, in order to generate the detailed Sgr core physics, including kinematics, and the way stars and dark matter have been stripped. 

This paper is structured as it follows. In Section 2, we focus on the modeling details. In Section 3 we discuss how the core properties are reproduced, and we also show the dark matter stripping history, then we show the stellar or dark matter distribution at the present time. The last Section includes the discussion and then the conclusions.

\begin{table*}
\centering
\caption{Initial condition parameters in this high resolution dynamical simulation.} 
\label{table1}
\centering
\begin{tabular}{lllll}
\hline
Parameter
         &
Milky Way    &
Sagittarius        &
Units             \\
\hline
Particle mass (star/dark matter)         &1.375(5.575)$\times10^{5}$  & 13.75(55.13) & M$_{\odot}$\\
Dark Matter mass     & 4.8$\times10^{11}$  & 9.0$\times10^{8}$  & M$_{\odot}$\\
Dark Matter scale     & 9.1 &1.6 & kpc\\
Stellar disc mass  & 3.58$\times10^{10}$    &   3.0$\times10^{7}$   & M$_{\odot}$\\
Stellar scale length & 2.4    & 0.3   & kpc\\
Stellar scale height & 0.24   & 0.15 & kpc\\
Bulge mass& 1.12$\times10^{10} $   & N/A  & M$_{\odot}$\\
Bulge scale & 0.4   & N/A   & kpc\\
Number of particle & 1.2  & 18    & Million\\

Initial Position &   &(66, -9, 27)& \,kpc \\
\hline
Initial Velocity &   &(-48, -17, 65)& \,km s$^{-1}$  \\
\hline
\end{tabular}
\end{table*}

\section{Modeling parameters}

Here we give a concise review of our high resolution modeling for which the mass resolution for dwarf galaxy is one hundred times better than the fiducial model in \citet{wang2022}, i.e., 13.75 solar mass per Sgr particle instead of 1375 solar mass . The Milky Way model has been constructed following \citet{Barnes2002}, and it includes a bulge, an exponential stellar disc, and a core dark matter halo, while the Sgr dwarf galaxy is made of a dark matter halo and of a disc (see Table~\ref{table1}). The halo density profile is converging at large radii.  More details about the modeling components could be found in \citet{Barnes2002} and in \citet{wang2022}. The modeling is realized using GIZMO \citep{Hopkins2015} and Table~\ref{table1} gives the initial conditions. In order to avoid numerical/artificial kick between heavy MW particles and light dwarf galaxy particles,  the technique of adaptive gravity softening (AGS, \citealt{Hopkins2018}) has been adopted. The minimal softening is set to 2.5 pc (e.g., for  interactions between light particles), which allows a full-resolution of dynamics within the core scalelength (300 pc) of Sgr dwarf galaxy. 

However, the adaptive softening is not a magic fix for having very different particle masses (see Table~\ref{table1}). We have first verified the absence of anomalous motions especially during the pericenter passages. Second, we have replaced the Milky Way (bulge, disk, and halo) particles by an analytic potential having precisely the same initial parameters (sizes and masses), and have verified whether the velocity dispersion of stars has been affected by numerical heating. The latter is detected only during the short-time scale passages to the pericenter, after which values of our model cannot be distinguished from that of the analytic potential. Third, we have ran a simulation with a higher resolution, in which the mass of MW particles has been divided by 10. We retrieve the same results, i.e., numerical heating mildly affects stellar velocity for short time, during pericenter passages. We then conclude that our simulations can be well representative of the Sgr core kinematics and morphology (See Appendix Figure~\ref{numericalheating}).

In this work the core/stream is projected in a  right-handed Galactocentric cartesian coordinates \citep[X, Y, and Z]{vanderMarel2002}. We have adopted the solar motion from \citet{2018RNAAS...2..210D} with [$U_{\odot}$ $V_{\odot}$ $W_{\odot}$] = [12.9, 12.6, 7.78] km s$^{-1} $. The circular speed of the Local Standard of Rest is adopted to be 233 km s$^{-1} $ \citep{2018RNAAS...2..210D}. The distance of the Sun from the Galactic center is chosen to be $R_{\odot}$  = 8.277 \,kpc \citep{2022A&A...657L..12G} and $Z_{\odot}$ = 20.8 \,pc \citep{2019MNRAS.482.1417B}, and we have verified that different solar motions and coordinates would not change our final conclusions. We have also verified that our models remain stable when evolved in isolation for several Gyr.

 \begin{figure*}
  \centering
  \includegraphics[width=1.0\textwidth]{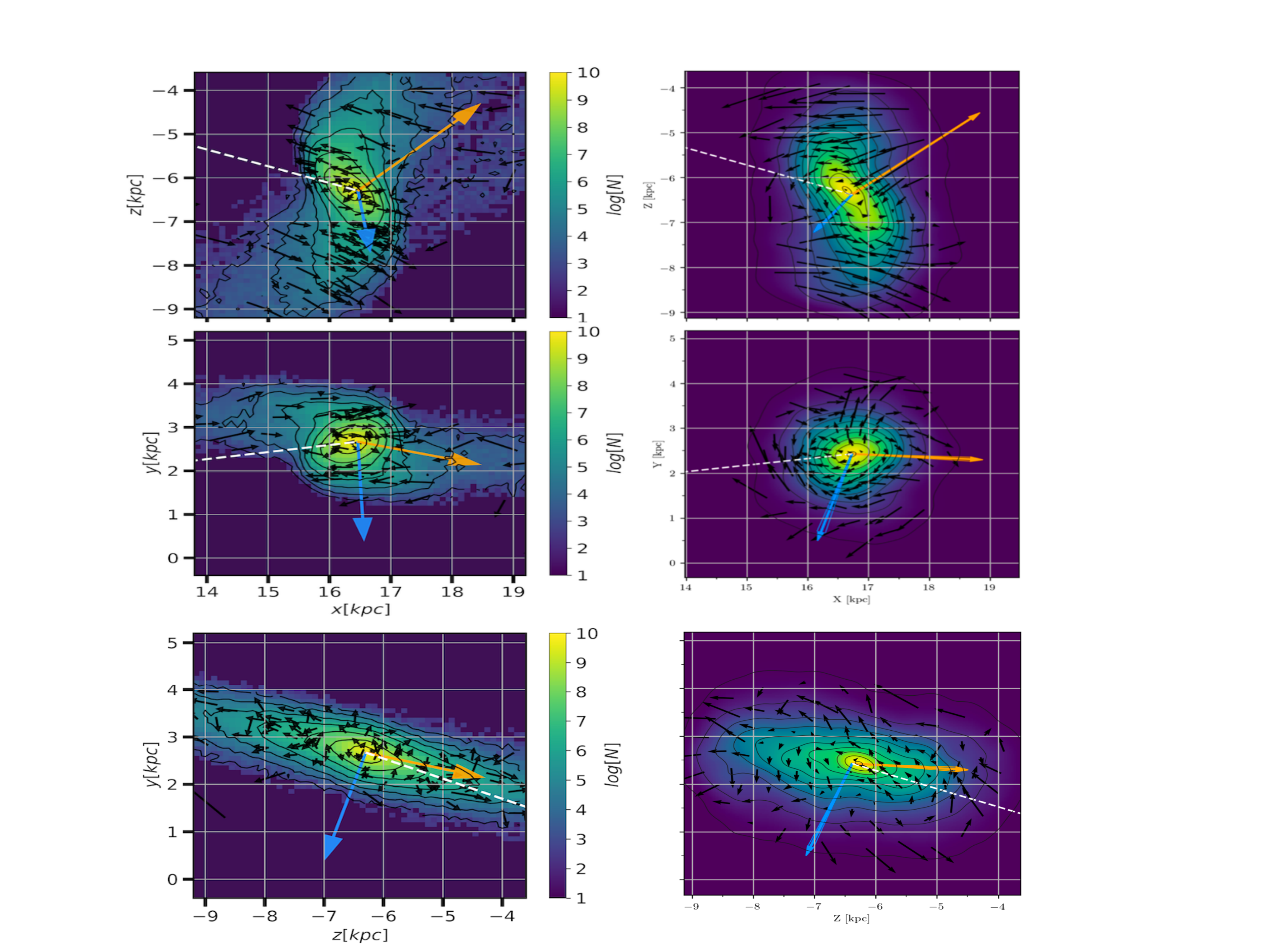}
  \caption{Left column: Sgr core density and kinematics of the model projected on the Galactocentric x-z, x-y, and z-y planes. Stellar density is shown by black contours and coloured from yellow to blue. The projected tangential velocities after subtraction of the COM velocity are shown by the black arrows and the orange arrows indicate the projected bulk velocity of the COM region. The blue arrow shows the projection of the internal angular momentum. The direction towards the MW center is indicated by the white dashed line. All arrows are scaled in order to be clearly shown for physical discussion. Right column: The observational results from \citet[see, their Fig. 10]{delPino2021} that are shown for comparison. A more detailed view of the stellar motions is shown in the Appendix Figure~\ref{corecountsnew}.}
  \label{corecounts}
\end{figure*}

\section{Results}

Figure~\ref{corecounts} shows the 3D stellar density and 3D kinematics distributions of the Sgr core for both modelling (left panels) and observations (right panels) from \citet[see,  their Fig. 10]{delPino2021}. The top panels reveal the S-shape in the x-z plane, which is surrounded by star motions indicated by black arrows that indicate stellar motions  (v$_{x}$, v$_{z}$) after subtraction of the COM (centre of mass) velocity. It suggests that tidal stripping becomes prominent at 2 kpc from the Sgr center, where the stars of the Sgr core are progressively removed from the main body to feed the stream. By accounting for this mechanism, the simulations reproduce well the observed morphology and kinematics of the Sgr core in the x-z plane.  Similarly, simulations shown in the middle left panel (x-y plane) present an almost circular shape as it is observed, as well as a relatively chaotic distribution of velocities shown as black arrows based on v$_{x}$, v$_{y}$. Black arrows in the z-y plane indicate an expansion of the stars in the outskirts of the core for both data and modeling (see bottom panels). Besides this, in the z-y plane the morphology of the modeled Sgr core appears to be more elongated than the observed one. 

We also note that the S-shape is caused by the later tidal disruption and it is not a long lived structure. A more detailed map of the Sgr core stellar motions is given in the Appendix (Figure~\ref{corecountsnew}).

In Figure~\ref{corecounts} the large blue arrow shows the projection of the internal angular momentum, for which a good agreement is found between observations and modeling, except perhaps in the x-z plane (top panel) for which their orientations differ by $\sim$ 30 degrees. The large orange arrow represents the COM velocity direction, for which observations and modeling agrees within 5 degrees. In the x-z  and y-z planes  the core shows a counter-clockwise and clockwise rotation, respectively. It appears that the model reproducing well the whole Sgr stream compares quite well with the observations of the Sgr core of \citet{delPino2021} for both morphology and kinematics. However the dynamics, and in particular for the internal angular momentum direction, is not fully recovered, and to reach a perfect matching between modeling and observations would require some further fine tuning. 

However, we think that the most important change to be done would be to introduce the gas in the Sgr progenitor, since it is likely a gas-rich irregular dwarf. Gas stripping history was revealed by \citet{Thor2018}, showing that the gas has been lost about 1\,Gyr ago. If adding gas could not improve the modeling, one may consider to increase the number of parameters, e.g., by adding an extra massive LMC or considering a non-spherical Milky Way halo, For example, \citet{Vasiliev2021} have added a massive LMC and a twisted halo, which helped them in reproducing the Sgr stream.

\begin{figure*}
  \centering
  \includegraphics[width=1.0\textwidth]{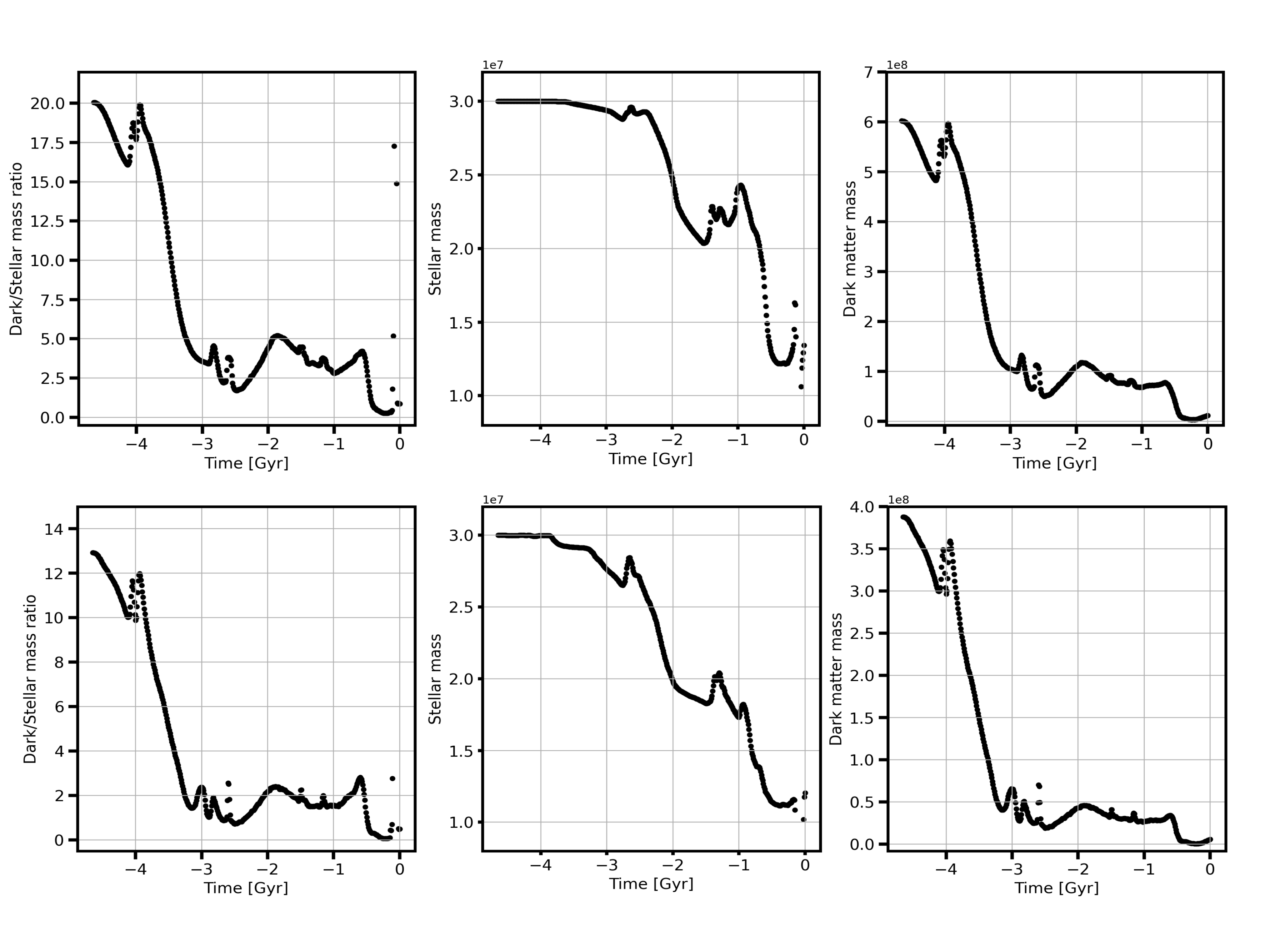}
  \caption{It shows the dark matter and stellar mass evolution history during our simulation over a 4.7\,Gyr duration (top panels: within a radius of 5 kpc; bottom panels: within a 2 kpc radius). The left, middle, and right panels show the ratio of dark matter to the stellar mass, the stellar stripping history of the core, and that of the dark matter.}
  \label{darkevolution}
\end{figure*}

Our modeling suggests a current Sgr core deficient in dark matter as it is shown in Figure~\ref{darkevolution}, which gives the time evolution of both dark and stellar matter within a radius of 5 \,kpc (top panels) and 2 kpc (bottom panels). The left panels  show the evolution of the  dark matter/stellar mass ratio that varies from an initial value of 20 (13 in the bottom panel) to almost 0 after 4.7 \,Gyr evolution. During the same elapsed time, the stellar mass decreases from 3$\times10^{7}$ M$_{\odot}$ to  1.4 (1.1)$\times10^{7}$ M$_{\odot}$, corresponding to a mass loss rate of 53\% (63\%) as seen in the middle panel of Figure~\ref{darkevolution}. This stellar mass loss contrasts with that of the dark matter, for which the  mass is decreasing from 6 (4)$\times10^{8}$ M$_{\odot}$ to almost 0 according to the right panels of Figure~\ref{darkevolution}. In other words,  almost all the dark matter content of Sgr has been lost from its core, and Sgr becomes almost dark matter free at the present epoch.

We also note that the model of \citet{Vasiliev2020} leads to a smaller elapsed time for the Sgr disruption, while also fitting the stream very well. This mainly comes from their MW model that is more massive than ours, leading to larger gravitational forces and then shorter times. Other differences are due to different choices in implementing the Sgr progenitor (e.g., disk or a spheroid, intial abundance of dark matter).

Moreover, it appears that dark matter losses occurred mostly during pericenter passages, which suffices to almost fully empty it,  3\,Gyr ago. Dark matter studies of dwarf galaxies have often excluded Sgr because of its strong interaction with the Milky Way. Such guesses are robustly confirmed by our study that suggests a dark matter deficient Sgr core, for which most properties are led by the Milky Way tidal forces.

\begin{figure}
  \centering
  \includegraphics[width=0.45\textwidth]{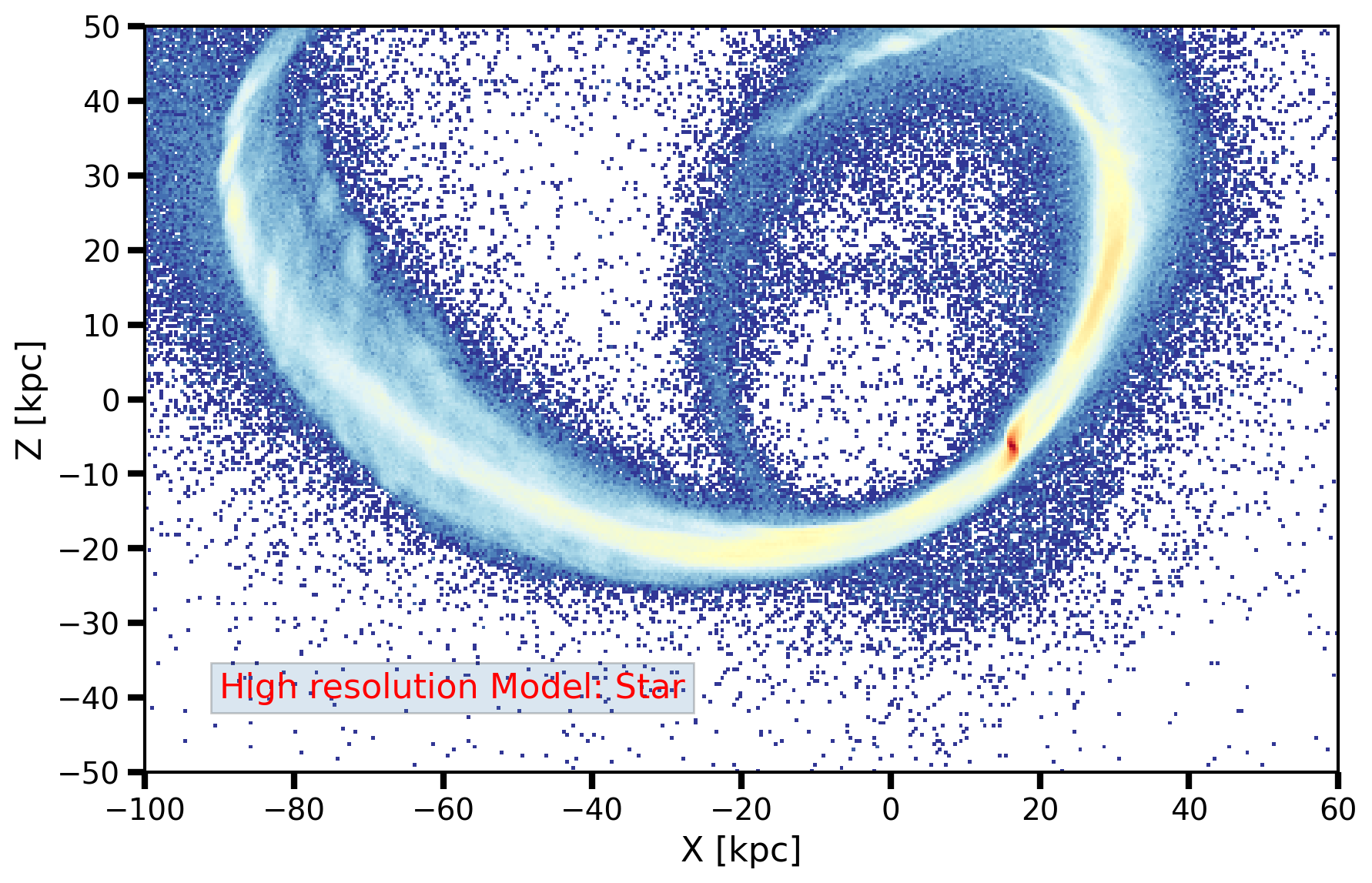}
   \centering
  \includegraphics[width=0.45\textwidth]{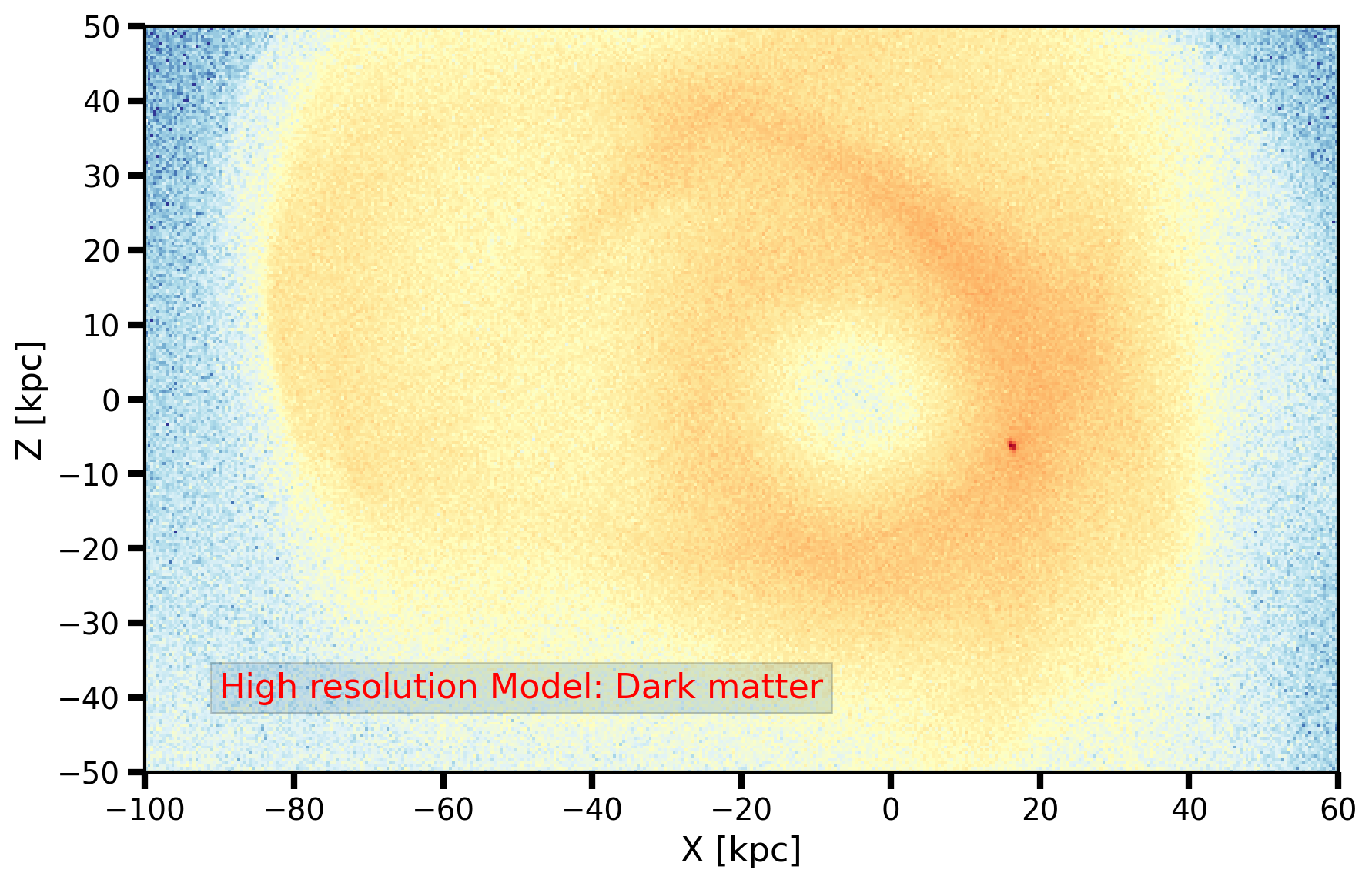}
  \caption{It shows stellar and dark matter tracks at the present time after 4.7\,Gyr evolution. The top panel shows the stellar leading, trailing arm, core, north and south bifurcation. The bottom panel indicates how the dark matter may distribute differently than stars.}
  \label{darktoday}
\end{figure}

Figure~\ref{darktoday} shows the leading, trailing, north and south bifurcation of the Sgr stellar stream in the (X, Z) plane. It also evidences how the distribution of dark matter particles (bottom panel) differs from that of the stars forming the Sgr stream (top panel). This illustrates well that dark matter populating galactic halo do not interact through gravitational torques and let the stream be almost only populated by stars.

\section{Discussion \& Conclusions}

In their modeling \citet{Vasiliev2020} found a remnant stellar mass within a 5\,kpc radius of $10^{8}$ M$_{\odot}$ assuming few kpc for the dark matter scale radius. In this work we give the evolution history of both dark matter and stellar masses within 2 or 5\,kpc in Figure~\ref{darkevolution} while the initial dark matter scale length is 1.6\,kpc. This contrasts with our finding of only 1.4 $\times10^{7}$ M$_{\odot}$ within the same radius. Determining the stellar mass of the Sgr core is not a simple exercise, because (i) it lies near the MW disk leading to severe extinction and confusion effects, and (ii) the core limits are not easy to determine as it is directly linked to the stream (see Figure~\ref{corecounts}) as expected from the Sgr tidally disrupted nature. For example, \citet[see also \citealt{Simon2019}]{Majewski2003} found that Sgr has a V-band luminosity of 2.1 $\times10^{7}$ L$_{\odot}$, which contrast with the much larger value found by \citet{Niederste2010}, who made an intensive stellar count study and find around 10 (9.6$-$13.2) $\times10^{7}$ L$_{\odot}$ for which 30\% is associated to the the Sgr core and 70\% to the stream. Assuming a factor 1.5 for converting luminosity into stellar mass for a galaxy containing intermediate-age stars, one can find 3 $\times10^{7}$ M$_{\odot}$ from \citealt{Simon2019}, and 4.5 $\times10^{7}$ M$_{\odot}$ for the core mass of Sgr from \citet{Niederste2010}, which is likely an upper limit since their estimates includes stars lying up to 10 kpc (see their Figure 9) instead of the 5 kpc radius adopted by \citet{Vasiliev2020}.
Given the above observational limits, it means that the core mass ($10^{8}$ M$_{\odot}$) recovered from the model of \citet{Vasiliev2020} may overestimate while this paper (1.4 $\times10^{7}$ M$_{\odot}$) may underestimate the core stellar mass by a factor between 2 to 3, respectively. Perhaps more important for our modelling is the good agreement that predicts the correct fraction of stars in the stream and in the core (see Figure~\ref{darkevolution}).

In summary, the model presented by \citet{wang2022} that reproduces the Sgr stream including its bifurcation behavior is also able to successfully reproduce most of the Sgr core properties as it is shown in this work. Without specific fine-tuning, the modeling reproduces most of the core morphology, including the S-shape, as well as most of its kinematics, including the tidally removed stars at its S-shape edges, which are forming the Sgr stream. We are however aware that there are some discrepancies between the modeling and the observations, and these are mostly linked to the internal angular momentum poles, which may suggest to investigate an initially gas-rich Sgr. 

As another result, this study shows how stars and dark matter are tidally removed from the core, the dark matter being almost completely stripped as it is expected from theory. This model also predicts that the Sgr core will disappear within 2\,Gyr, letting only its central cluster intact\footnote{This can be seen in a video showing the evolution of star particles as shown in the top of Figure~\ref{darktoday}: \url{https://youtu.be/EqjEQpWaeIU}.}.  

\section*{Acknowledgments}
The author gratefully acknowledges the support of K.C. Wong Education Foundation. We appreciate the support of the International Research Program Tianguan, which is an agreement between the CNRS in France, NAOC, IHEP, and the Yunnan Univ. in China. YY thanks the National Natural Foundation of China (NSFC No. 11973042). Simulations in this  work  were performed at the  High-performent calculation (HPC)  resources MesoPSL financed by the project Equip@Meso (reference ANR-10-EQPX-29-01) of the program "Investissements d'Avenir" supervised by the 'Agence Nationale de la Recherche'. We are very grateful to Gary Mamon, Eugene Vasiliev, Frederic Arenou, Carine Babusiaux, Piercarlo Bonifacio, Jiang Chang, Yongjun Jiao and Marcel Pawlowski for the numerous and illuminating discussions we have had on the Sgr simulation. We thank the referee for giving us precious advices and suggestions that have seriously improved the manuscript.  HFW acknowledge the science research grants from the China Manned Space Project with NO. CMS-CSST-2021-B03. We are grateful to Phil Hopkins who kindly shared with us the access to the Gizmo code.

\section*{DATA AVAILABILITY}
The data underlying this article will be shared on reasonable request to the corresponding author.

\appendix

\section{More comparisons and investigations}
 \begin{figure*}
  \centering
  \includegraphics[width=1.0\textwidth]{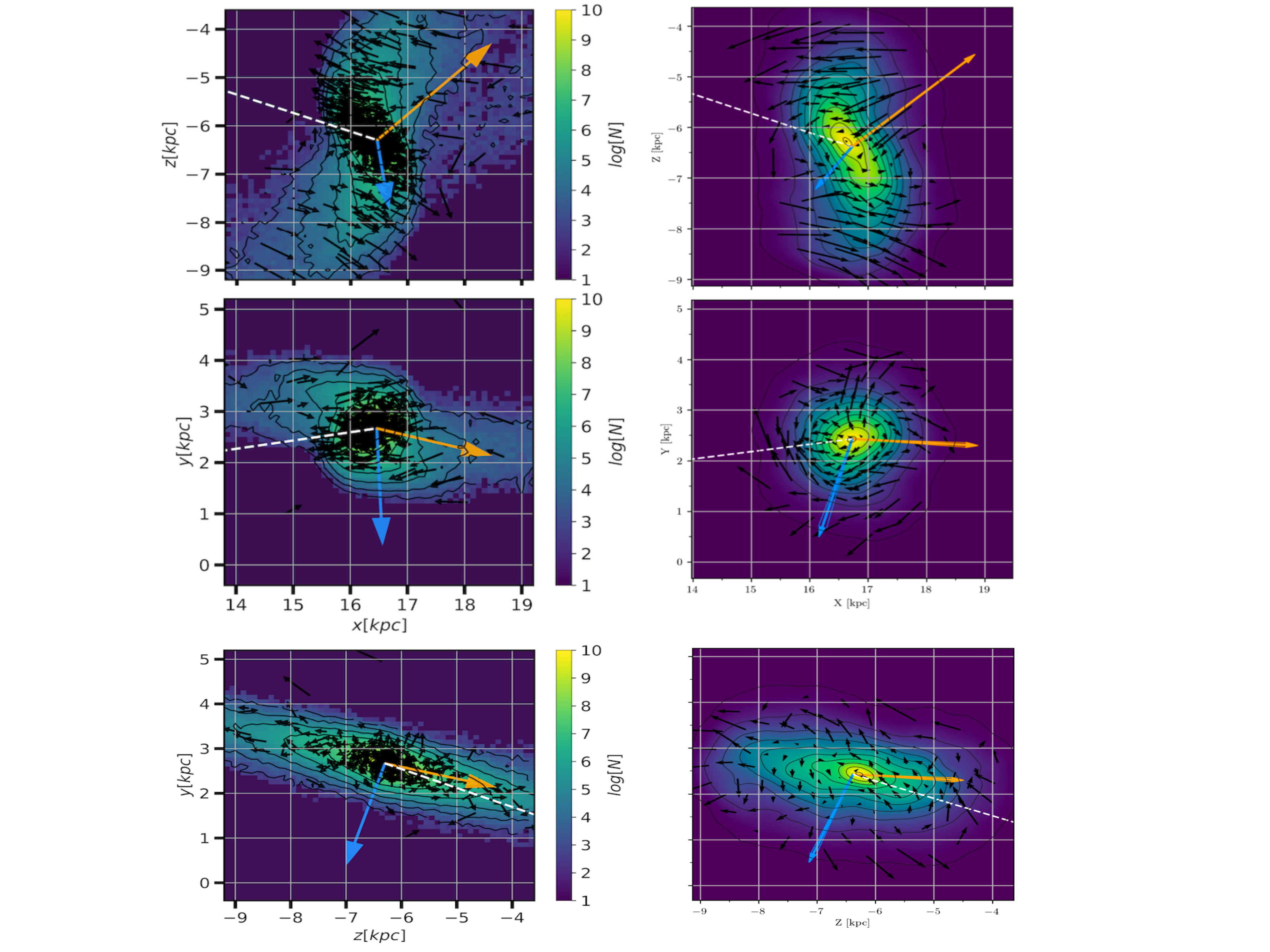}
  \caption{Similar to Figure~\ref{corecounts}, but with more arrows and points to make the rotation and expansion in the outskirts clearer. Left column: Sgr core density and kinematics of the model projected on the Galactic centric x-z, x-y, and z-y planes. Right column: the observational results from \citet[see, their Fig. 10]{delPino2021}that are shown for more comparison.}
  \label{corecountsnew}
\end{figure*}

 \begin{figure*}
 \centering
 \includegraphics[width=0.8\textwidth]{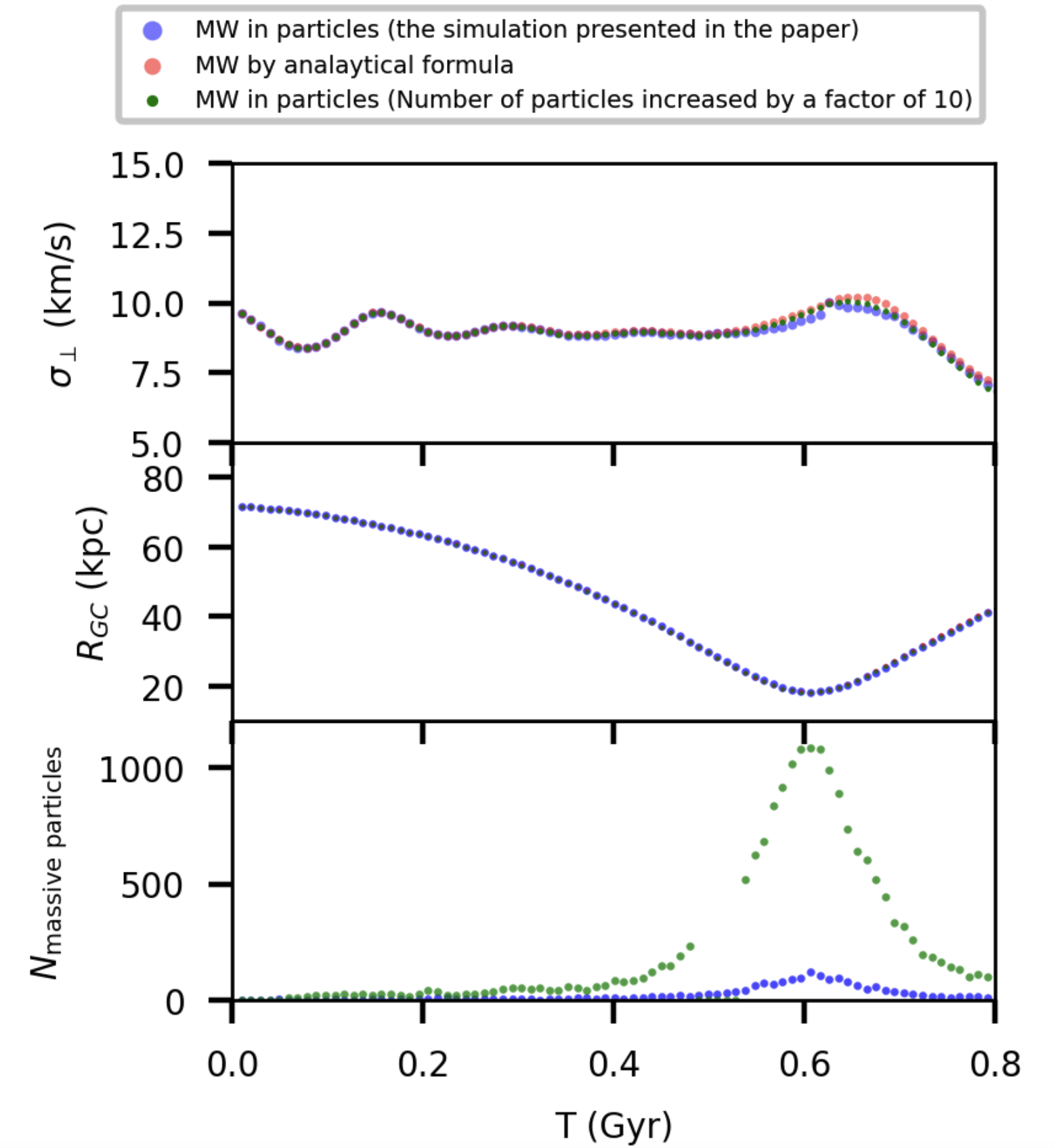}
 \caption{It compares our full-particle simulation presented in the paper and this simulation with the analytical MW (see the blue and red dots respectively). The latter is assumed without numerical heating, since it does not possess massive particles. From top to bottom: the velocity dispersion perpendicular to the orbital plane, as it is a quantity that could be only affected by the Sgr disk particle motions; the distance of dwarf galaxy from the MW center; the number of massive MW particles within the Sgr core (with a radius of 2 kpc). For a double check, we ran a new simulation by increasing the resolution of MW by a factor of 10, see the green dots. It shows a very smooth evolution in the velocity dispersion perpendicular to the orbital plane of dwarf galaxy, which verifies the concern that using less massive MW particles reduces the artificial kicking. These tests confirm that the artificial kicking happens only for short durations near the pericenter passage, and has a marginal impact to the results and conclusions of the paper.}
 \label{numericalheating}
\end{figure*}


\begin{thebibliography}{}
\bibitem[Antoja et al.(2018)]{Antoja2018}Antoja, T., Helmi, A., Romero-G\'{o}mez, M., et al. 2018, Nature, 561, 360
\bibitem[Barnes et al.(2002)]{Barnes2002}Barnes J. E., 2002, MNRAS, 333, 481
\bibitem[Bellazzini et al.(2020)]{Bellazzini2020}Bellazzini, M., Ibata, R., Malhan, K., et al., 2020, A\&A, 636, A107
\bibitem[Bland-Hawthorn et al.(2021)]{Joss2021}Bland-Hawthorn J., Tepper-Garc\'{i}a T., 2021, MNRAS, 504, 3168
\bibitem[Bennett \& Bovy(2019)]{2019MNRAS.482.1417B} Bennett, M. \& Bovy, J.\ 2019, \mnras, 482, 1417. doi:10.1093/mnras/sty2813
\bibitem[Belokurov  et al.(2006)]{Belokurov2006}Belokurov V., Zucker D., Evans N.W., et al., 2006, ApJ, 642, L137
\bibitem[Dehnen et al.(1993)]{Dehnen1993}Dehnen W., 1993, MNRAS, 265, 250
\bibitem[Drimmel \& Poggio(2018)]{2018RNAAS...2..210D} Drimmel, R. \& Poggio, E.\ 2018, Research Notes of the American Astronomical Society, 2, 210. doi:10.3847/2515-5172/aaef8b
\bibitem[\protect\citeauthoryear{del Pino et al.}{2021}]{delPino2021}del Pino A., Fardal M. A., van der Marel R. P., et al., 2021, ApJ, 908, 244
\bibitem[GRAVITY Collaboration et al.(2022)]{2022A&A...657L..12G} GRAVITY Collaboration, Abuter, R., Aimar, N., et al.\ 2022, \aap, 657, L12. doi:10.1051/0004-6361/202142465
\bibitem[Hammer et al.(2005)]{Hammer2005} Hammer, F., Flores, H., Elbaz, D., et al.\ 2005, \aap, 430, 115
\bibitem[Hammer et al.(2009)]{Hammer2009} Hammer, F., Flores, H., Puech, M., et al.\ 2009, \aap, 507, 1313
\bibitem[Hammer et al.(2020)]{Hammer2020} Hammer, F., Yang, Y., Arenou, F., et al.\ 2020, \apj, 892, 3
\bibitem[Hammer et al.(2021)]{Hammer2021} Hammer, F., Wang, J., Pawlowski, M.~S., et al.\ 2021, \apj, 922, 93
\bibitem[\protect\citeauthoryear{Hopkins et al.}{2015}]{Hopkins2015}Hopkins P. F., 2015, MNRAS, 450, 53
\bibitem[\protect\citeauthoryear{Hopkins et al.}{2018}]{Hopkins2018}Hopkins P. F., et al., 2018b, MNRAS, 480, 800
\bibitem[\protect\citeauthoryear{Hernquist et al.}{1993}]{Hernquist1993}Hernquist L., 1993, Survey. ApJ Suppl. Ser., 86, 389
\bibitem[\protect\citeauthoryear{Ibata et al.}{1994}]{Ibata1994}Ibata R., Gilmore G., Irwin M., 1994, Nature, 370, 194
\bibitem[\protect\citeauthoryear{Ibata et al.}{1995}]{Ibata1995}Ibata R., Gilmore G., Irwin M., 1995, MNRAS, 277, 781
\bibitem[Koposov et al.(2012)]{2012ApJ...750...80K} Koposov, S.~E., Belokurov, V., Evans, N.~W., et al.\ 2012, \apj, 750, 80. doi:10.1088/0004-637X/750/1/80
\bibitem[\protect\citeauthoryear{{\L}okas et al.}{2010}]{Lokas2010}{\L}okas, E. L., Kazantzidis, S., Majewski, S. R., et al. 2010, ApJ, 725, 1516
\bibitem[L\'{o}pez-Corredoira et al.(2020)]{Lopez2020} L\'{o}pez-Corredoira, M., Garz\'{o}n, F., Wang, H.-F., et al. 2020, \ A$\&$A, 634, A66
\bibitem[Laporte et al.(2018)]{2018MNRAS.481..286L} Laporte, C.~F.~P., Johnston, K.~V., G{\'o}mez, F.~A., et al.\ 2018, \mnras, 481, 286. doi:10.1093/mnras/sty1574
\bibitem[Laporte et al.(2019)]{2019MNRAS.485.3134L} Laporte, C.~F.~P., Minchev, I., Johnston, K.~V., et al.\ 2019, \mnras, 485, 3134. doi:10.1093/mnras/stz583
\bibitem[Laporte et al.(2020)]{2020MNRAS.492L..61L} Laporte, C.~F.~P., Belokurov, V., Koposov, S.~E., et al.\ 2020, \mnras, 492, L61. doi:10.1093/mnrasl/slz167
\bibitem[\protect\citeauthoryear{Mart\'{i}nez-Garc\'{i}a et al.}{2021}]{Martinez-Garcia2021}Mart\'{i}nez-Garc\'{i}a, A. M., del Pino, A., Aparicio, A., et al., 2021, MNRAS, 505, 5884
\bibitem[\protect\citeauthoryear{Navarro et al.}{1995}]{Navarro1995}Navarro J. F., Frenk C. S., White S. D. M., 1995, MNRAS, 275, 720
\bibitem[Majewski et al.(2003)]{Majewski2003} Majewski, S.~R., Skrutskie, M.~F., Weinberg, M.~D., et al.\ 2003, \apj, 599, 1082. doi:10.1086/379504
\bibitem[Niederste-Ostholt et al.(2010)]{Niederste2010} Niederste-Ostholt, M., Belokurov, V., Evans, N.~W., et al.\ 2010, \apj, 712, 516. doi:10.1088/0004-637X/712/1/516
\bibitem[Oria et al.(2022)]{2022ApJ...932L..14O} Oria, P.-A., Ibata, R., Ramos, P., et al.\ 2022, \apjl, 932, L14. doi:10.3847/2041-8213/ac738c
\bibitem[\protect\citeauthoryear{Penarrubia et al.}{2010}]{Penarrubia2010}Penarrubia J., Belokurov V., Evans N. W., et al., 2010, MNRAS, 408, L26
\bibitem[\protect\citeauthoryear{Penarrubia et al.}{2011}]{Penarrubia2011}Penarrubia J., Zucker D., Irwin M., et al., 2011, ApJ, 727, L2
\bibitem[Reid et al.(2014)]{Reid14}Reid, M. J., Menten, K. M., Brunthaler, A., et al. 2014, ApJ, 783, 130
\bibitem[Robertson \& Bullock(2008)]{Robertson2008} Robertson, B.~E. \& Bullock, J.~S.\ 2008, \apjl, 685, L27
\bibitem[Simon(2019)]{Simon2019} Simon, J.~D.\ 2019, \araa, 57, 375. doi:10.1146/annurev-astro-091918-104453\bibitem[Stewart et al.(2009)]{Stewart2009} Stewart, K.~R., Bullock, J.~S., Wechsler, R.~H., et al.\ 2009, \apj, 702, 307
\bibitem[Tepper-Garc\'{i}a  \& Bland-Hawthorn et al.(2018)]{Thor2018}Tepper-Garc\'{i}a, T., \& Bland-Hawthorn, J. 2018, MNRAS, 478, 5263
\bibitem[\protect\citeauthoryear{Vasiliev et al.}{2020}]{Vasiliev2020}Vasiliev E., Belokurov V., 2020, MNRAS, 497, 4162 (VB20)
\bibitem[\protect\citeauthoryear{Vasiliev et al.}{2021}]{Vasiliev2021}Vasiliev, E., Belokurov, V., \& Erkal, D. 2021, MNRAS, 501, 2279
\bibitem[\protect\citeauthoryear{van der Marel et al.}{2002}]{vanderMarel2002}van der Marel, R. P., Alves, D. R., Hardy, E., \& Suntzeff, N. B. 2002, AJ, 124, 2639
\bibitem[\protect\citeauthoryear{White \& Rees}{1978}]{White1978}White S. D. M., Rees M. J., 1978, MNRAS, 183, 341
\bibitem[Wang et al.(2018a)]{wang2018a} Wang, H.~F., L\'{o}pez-Corredoira, M., Carlin, J. L., {et~al.,} 2018a, MNRAS, 477, 2858
\bibitem[Wang et al.(2018b)]{wang2018b}  Wang, H. F., Liu, C., Xu, Y., et al., 2018b, MNRAS, 478, 3367
\bibitem[Wang et al.(2019)]{wang2019} Wang, H. F., Carlin, J. L., Huang, Y., et al. 2019, ApJ, 884, 135
\bibitem[Wang et al.(2020a)]{wang2020a} Wang, H.~F., L\'{o}pez-Corredoira, M., Huang, Y., {et~al.,} 2020a, MNRAS, 491, 2104 
\bibitem[Wang et al.(2020c)]{wang2020b} Wang, H.~F., L\'{o}pez-Corredoira, M., Huang, Y., et al., 2020b, ApJ, 897, 119
\bibitem[Wang et al.(2020d)]{wang2020c}  Wang, H. F., Huang, Y., Zhang, H.-W., et al., 2020c, ApJ, 902, 70
\bibitem[Wang et al.(2022)]{wang2022}  Wang, H. F., Yang, Y. B.,  Hammer, F., et al., 2022, eprint arXiv:2204.08542
\bibitem[Yu et al.(2021)]{Yu2021}Yu Y., Wang H.-F., Cui W.-Y., et al., 2021, ApJ, 922, 80 
\bibitem[Yang et al.(2022)]{Yang2022} Yang, P., Wang, H.-F., Luo, Z.-Q., et al.\ 2022, eprint arXiv:2205.09227

\end{thebibliography}
\end{document}